\documentclass{desyproc}

\usepackage{amsmath}
\usepackage{amssymb}

\begin{document}

\title{Next-to-leading order matrix elements\\ and truncated showers}

\author{{\slshape Stefan H{\"o}che$^1$\footnote{Speaker}, Frank Krauss$^{2,3}$,
  Marek Sch{\"o}nherr$^4$, Frank Siegert$^{2,5}$}\\[1ex]
  $^1$ Institut f{\"u}r Theoretische Physik, 
  Universit{\"a}t Z{\"u}rich, CH-8057 Zurich, Switzerland\\
  $^2$ Institute for Particle Physics Phenomenology,
  Durham University, Durham DH1 3LE, UK\\
  $^3$ PH-TH, CERN, CH-1211 Geneva 23, Switzerland\\
  $^4$ Institut f{\"u}r Kern- und Teilchenphysik,
  TU Dresden, D-01062, Dresden, Germany\\
  $^5$ Department of Physics \& Astronomy,
  University College London, London WC13 6BT, UK\\}

\contribID{xy}  
\confID{1964}
\desyproc{DESY-PROC-2010-01}
\acronym{PLHC2010}
\doi            

\maketitle

\begin{abstract}
  An algorithm is presented that combines the ME+PS approach to merge sequences 
  of tree-level matrix elements into inclusive event samples~\cite{Hoeche:2009rj} 
  with the POWHEG method, which combines exact next-to-leading order matrix elements 
  with parton showers~\cite{Nason:2004rx}. 
  The quality of the approach and its implementation in Sherpa~\cite{Gleisberg:2008ta}
  are exemplified by results for $e^+e^-$ annihilation into hadrons at LEP, 
  for Drell-Yan lepton-pair production at the Tevatron and 
  for Higgs-boson and $W^+W^-$-production at LHC energies.
\end{abstract}

\section{Introduction}
Facing the huge progress at the LHC, with first data taken, 
and first results already published, it is crucial to have reliable tools at hand 
for the full simulation of Standard Model signal and background processes as well as 
for the simulation of signals for new physics. This task is universally handled 
by Monte-Carlo event generators like Sherpa~\cite{Gleisberg:2008ta}.

One of the key features of such advanced Monte-Carlo programs is the possibility 
to consistently combine higher-order tree-level matrix element events with subsequent 
parton showers (ME+PS)~\cite{Hoeche:2009rj}. This feature has proved invaluable in 
various recent analyses of data from previous experiments, which are sensitive to 
large-multiplicity final states. Despite being a tremendous improvement over pure
leading-order theory, ME+PS merging 
still suffers from one major drawback of all tree-level calculations, which is their 
instability with respect to scale variations. This deficiency ultimately necessitates 
the implementation of NLO virtual corrections in Monte-Carlo programs. Two universally 
applicable methods were suggested in the past, which can perform this task, and whereof 
one is the so-called POWHEG algorithm~\cite{Nason:2004rx}. This technique has been 
reformulated in~\cite{Hoeche:2010pf}, such that it can be applied in an automated manner.

Having implementations of both, ME+PS merging and the POWHEG method at our disposal,
the question naturally arises, whether the two approaches can be combined into an even 
more powerful one, joining their respective strengths and eliminating their weaknesses. 
A first step into this direction was taken independently in~\cite{Hamilton:2010wh} 
and in~\cite{Hoeche:2010kg}. Here we will summarise the essence of the algorithms 
presented ibidem and exemplify the quality of related Monte-Carlo predictions.

\section{The MENLOPS approach}
A formalism allowing to describe both, the ME+PS and the POWHEG method on the same 
footing was introduced in~\cite{Hoeche:2010pf}. To compare, and, ultimately, to 
combine both methods, only the expressions for the differential cross section 
describing the first emission off a given core process must be worked out; this 
is where the combination takes place.

In a simplified form, the expectation value of an observable in the POWHEG method can be 
described by the following master formula (for details see~\cite{Nason:2004rx,Hoeche:2010pf})
\begin{equation}\label{Eq:master_powheg}
  \langle O\rangle^{\rm POW}=\sum_i\int{\rm d}\Phi_B\,\bar{\rm B}_i(\Phi_B)
  \Bigg[\; 
    \underbrace{\bar{\Delta}_i(t_0)\,O(\Phi_B)}_\text{no emission} + 
    \underbrace{\sum_j\int_{t_0} {\rm d}\Phi_{R|B}\,\frac{{\rm R}_j(\Phi_R)}{{\rm B}_i(\Phi_B)}\;
      \bar{\Delta}_i(t)\,O(\Phi_R)\;}_\text{resolved emission}
  \Bigg]\;,
\end{equation}
where $\bar{\rm B}_i(\Phi_B)$ is the NLO-weighted differential cross section for the 
Born phase-space configuration $\Phi_B$ and 
$\bar{\Delta}_i(t)=\exp\left\{\,-\sum_j\int_t{\rm d}\Phi_{R|B}\,\rm R_j/B_i\,\right\}$ 
is the so-called POWHEG-Sudakov form factor. The indices $i$ and $j$ label parton
configurations, see~\cite{Hoeche:2010pf}. The parameter $t$ is the ordering variable 
of the underlying parton-shower model and $t_0$ is the respective cutoff. Hence, $t$ is 
one of the variables used to parametrise the radiative phase space $\Phi_{R|B}$. 

In a similar manner, a simplified master formula for the expectation value of $O$ in the 
ME+PS approach can be derived. It reads (for details see~\cite{Hamilton:2010wh,Hoeche:2010kg})
\begin{equation}\label{Eq:master_meps}
  \begin{split}
  &\langle O\rangle^{\rm ME+PS}=\sum_i\int{\rm d}\Phi_B\,{\rm B}_i(\Phi_B)
  \Bigg[\; 
    \underbrace{\Delta_i(t_0)\,O(\Phi_B)}_\text{no emission} + 
    \sum_j\int_{t_0}{\rm d}\Phi_{R|B}\\
    &\qquad\times\Big(
      \underbrace{\Theta(Q_{\rm cut}-Q)\,\frac{8\pi\alpha_s}{t}\,
        \mathcal{K}_{R_j|B_i}\,\frac{\mathcal{L}_{R_j}}{\mathcal{L}_{B_i}}}_{\text{PS domain}}+
      \underbrace{\Theta(Q-Q_{\rm cut})\,
        \frac{{\rm R}_j(\Phi_R)}{{\rm B}_i(\Phi_B)}}_{\text{ME domain}}
    \Big)\,\Delta_i(t)\,O(\Phi_R)
  \;\Bigg]\;.
  \end{split}
\end{equation}
The terms labelled ``ME domain'' and ``PS domain'' describe the probability of additional 
QCD radiation according to the real-radiation matrix elements and their corresponding 
parton-shower approximations, respectively. In this context, $\mathcal{K}_{R_j|B_i}$ are 
the parton-shower evolution kernels and $\mathcal{L}_{R_j/B_i}$ are the parton luminosities 
of the real-emission and the underlying Born configurations. In contrast to 
$\bar{\Delta}_i(t)$ in Eq.~\eqref{Eq:master_powheg}, $\Delta_i(t)$ is the uncorrected 
Sudakov form factor of the parton-shower model.

Combining the ME+PS method with POWHEG essentially amounts to combining the two above
equations into a new master formula for the MENLOPS approach. This expression reads
\begin{equation}\label{Eq:master_menlops}
  \begin{split}
  \langle O\rangle^{\rm MENLOPS}=&\sum_i\int{\rm d}\Phi_B\,\bar{\rm B}_i(\Phi_B)
  \Bigg[\; 
    \underbrace{\bar{\Delta}_i(t_0)\,O(\Phi_B)}_\text{no emission} + 
    \sum_j\int_{t_0}{\rm d}\Phi_{R|B}\,\frac{{\rm R}_j(\Phi_R)}{{\rm B}_i(\Phi_B)}\\
    &\qquad\times\Big(
      \underbrace{\Theta(Q_{\rm cut}-Q)\,\bar{\Delta}_i(t)}_{\text{PS domain}}+
      \underbrace{\Theta(Q-Q_{\rm cut})\,\Delta_i(t)}_{\text{ME domain}}
    \Big)\,O(\Phi_R)
  \;\Bigg]\;.
  \end{split}
\end{equation}
In order to restore the POWHEG master formula, the ``ME domain'' term would have to be
multiplied by the ratio of Sudakov form factors $\bar{\Delta}_i(t)/\Delta_i(t)$ only.
Expanding this ratio to first order reveals that the above formula automatically
yields next-to-leading order accurate predictions for any infrared and collinear safe
observable $O$.

\section{Results}
In the following we present selected results obtained with an implementation of the 
ME\-NLO\-PS algorithm in the Sherpa event generator.  In particular we aim at detailing 
the improved description of data collected in various collider experiments.

\begin{figure}[h]
  \begin{center}\vspace*{-2mm}
    \includegraphics[scale=0.55]{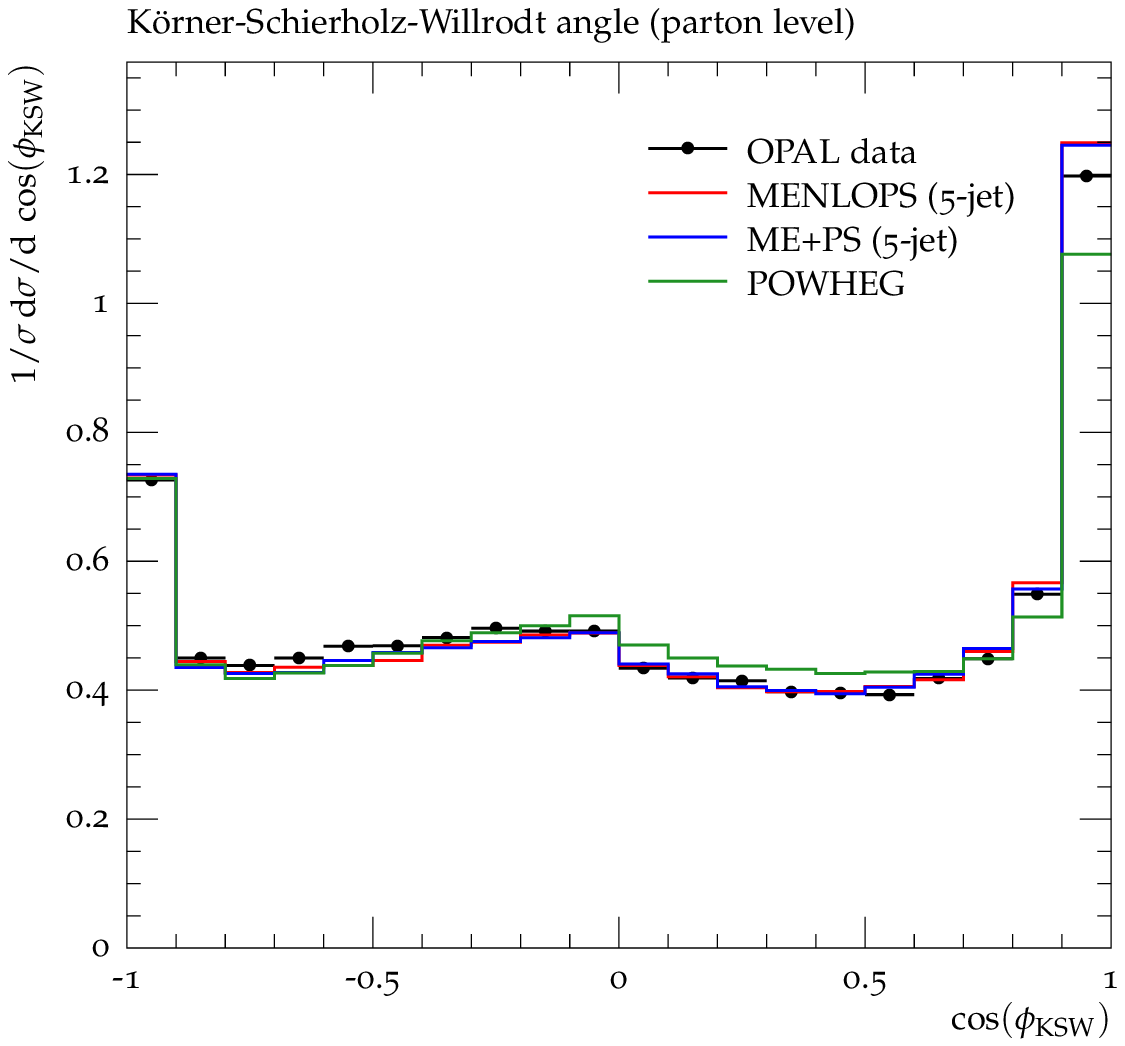}\hspace*{0.05\textwidth}
    \includegraphics[scale=0.55]{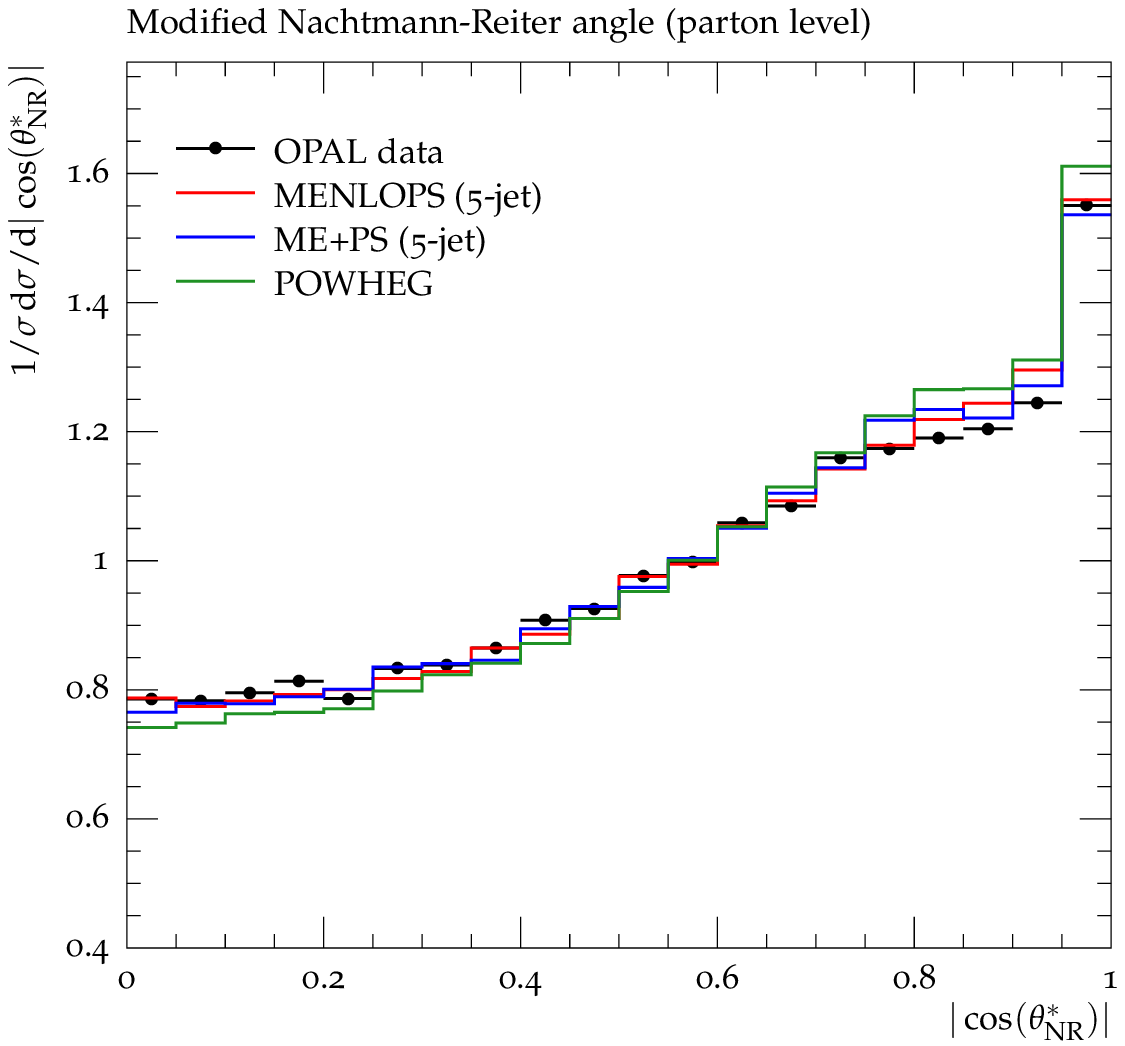}\vspace*{-5mm}
  \end{center}
  \caption{The K{\"o}rner-Schierholz-Willrodt (KSW) angle (left)
  and the modified Nachtmann-Reiter (NR) angle (right) in four-jet events 
  defined using the Durham algorithm with $y_\mathrm{cut}=0.008$. 
  Results at the parton level are compared to data from the 
  OPAL experiment~\cite{Abbiendi:2001qn}.
  \label{Fig:ee_fourjetangles}}
\end{figure}

We focus first on electron-positron annihilation into hadrons at LEP energies 
($\sqrt{s}=$91.25 GeV). Virtual matrix elements
were supplied by BlackHat~\cite{Berger:2009zg}. Figure~\ref{Fig:ee_fourjetangles} 
displays distributions of selected angular correlations in 4-jet production, 
that have been important for tests of perturbative QCD. The good fit to those data 
proves that correlations amongst the final-state partons are correctly implemented
by the higher-order matrix elements in the MENLOPS method.

\begin{figure}[h]
  \begin{center}\vspace*{-5mm}
    \includegraphics[scale=0.55]{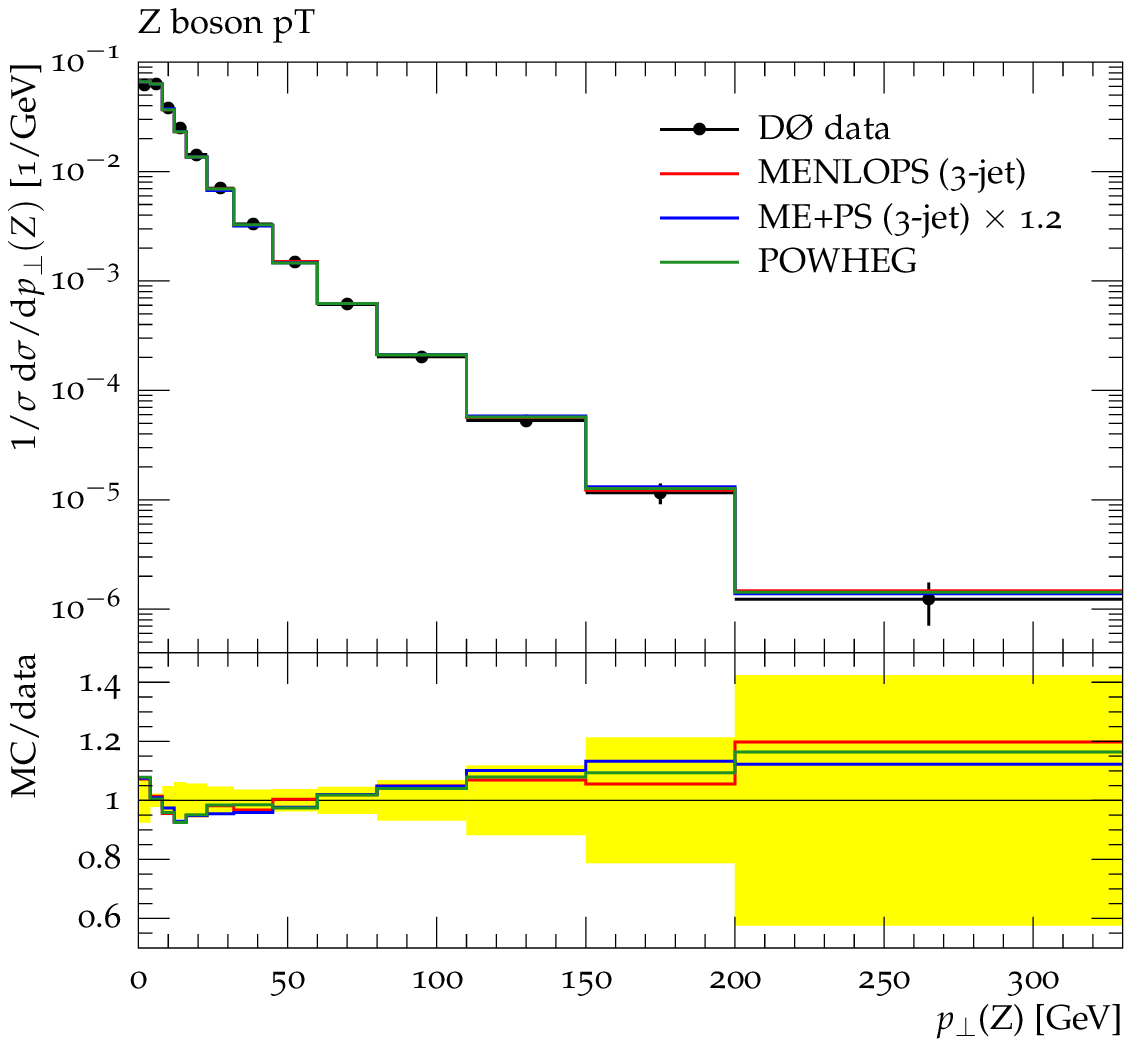}\hspace*{0.05\textwidth}
    \includegraphics[scale=0.55]{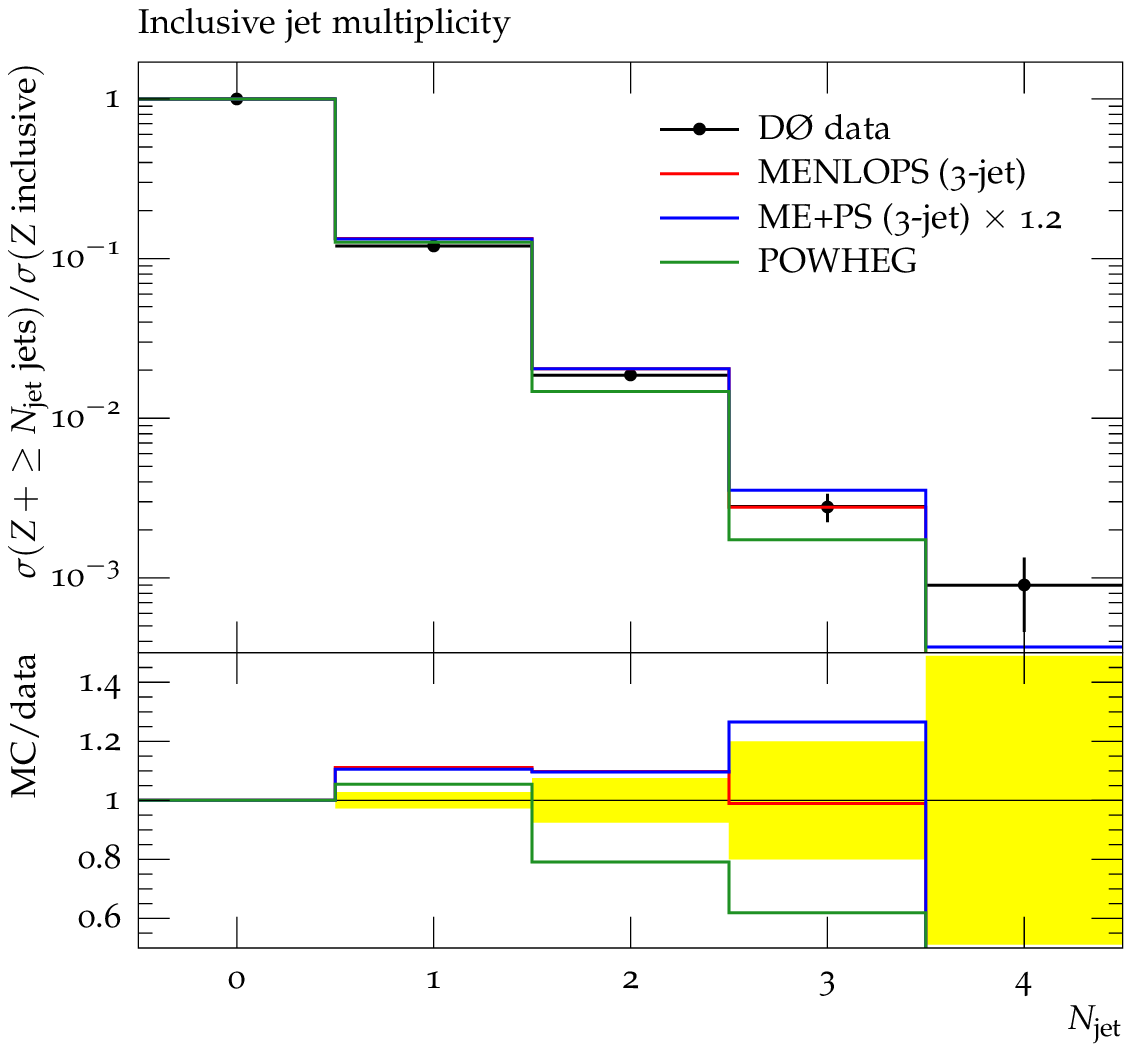}\vspace*{-5mm}
  \end{center}
  \caption{Left: The transverse momentum of the reconstructed $Z$ boson in Drell-Yan
  events at the Tevatron ($\sqrt{s}=1.96$ TeV). Results obtained with the
  MENLOPS approach are compared to data from the D\O\ experiment~\cite{Abazov:2010kn}.
  Right: Inclusive jet multiplicity in Drell-Yan
  events. Monte-Carlo predictions are compared to data from~\cite{Abazov:2006gs}.
  \label{Fig:dy_zpt_jetmulti}}
\end{figure}

Similar findings apply in the analysis of the Drell-Yan process at Tevatron energies
($\sqrt{s}=$1.96 TeV). Figure~\ref{Fig:dy_zpt_jetmulti} shows the transverse momentum 
distribution of the reconstructed $Z$-boson and the multiplicity distribution of 
accompanying jets, constructed using the D\O\ improved legacy cone algorithm with 
a cone radius of $R=0.5$, $p_{\perp,j}>20$~GeV and $|\eta_j|<2.5$. The agreement 
of the MENLOPS result with the respective data is outstanding.

\begin{figure}[p]
  \begin{center}
    \includegraphics[scale=0.5725]{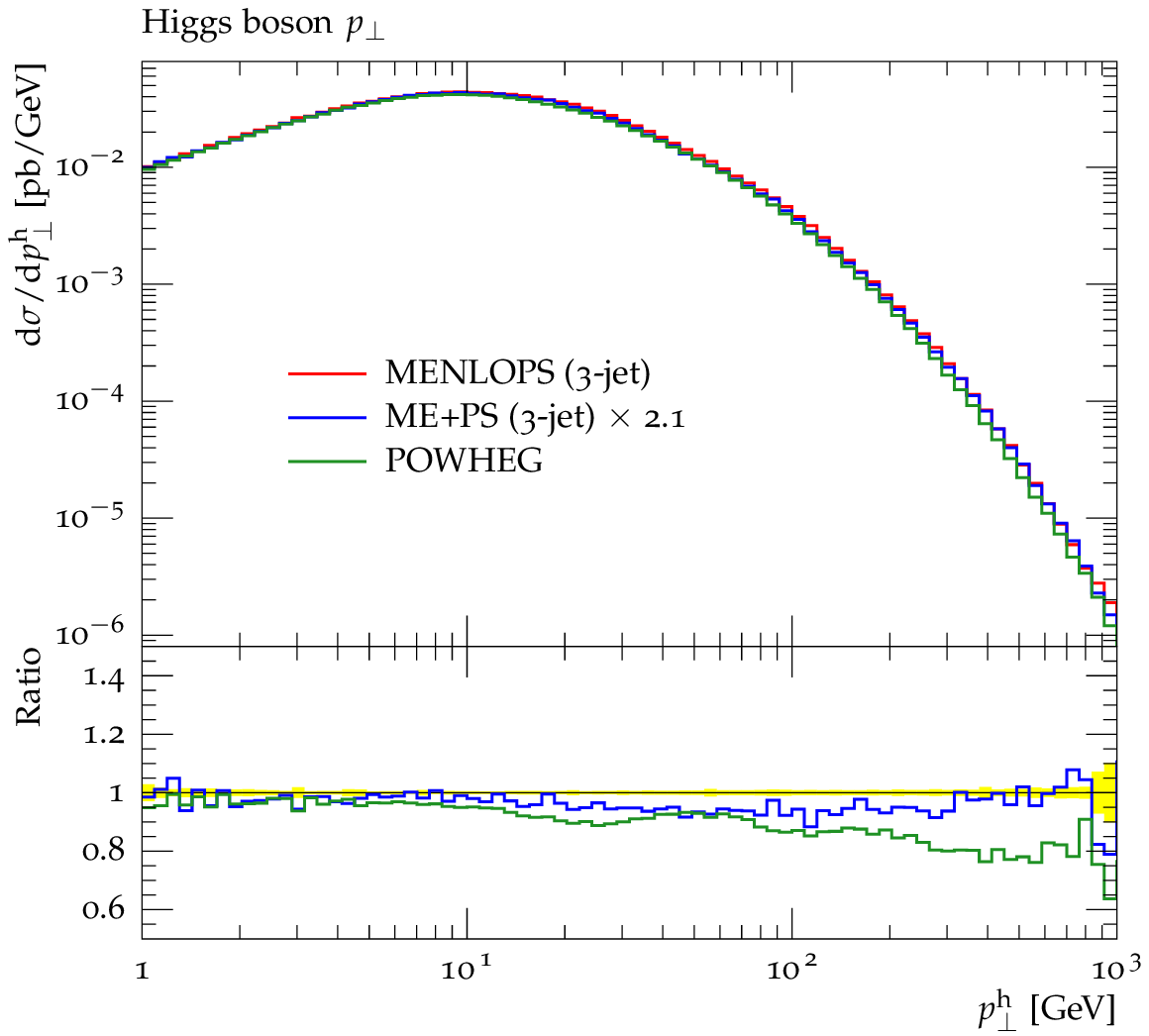}\hfill
    \includegraphics[scale=0.575]{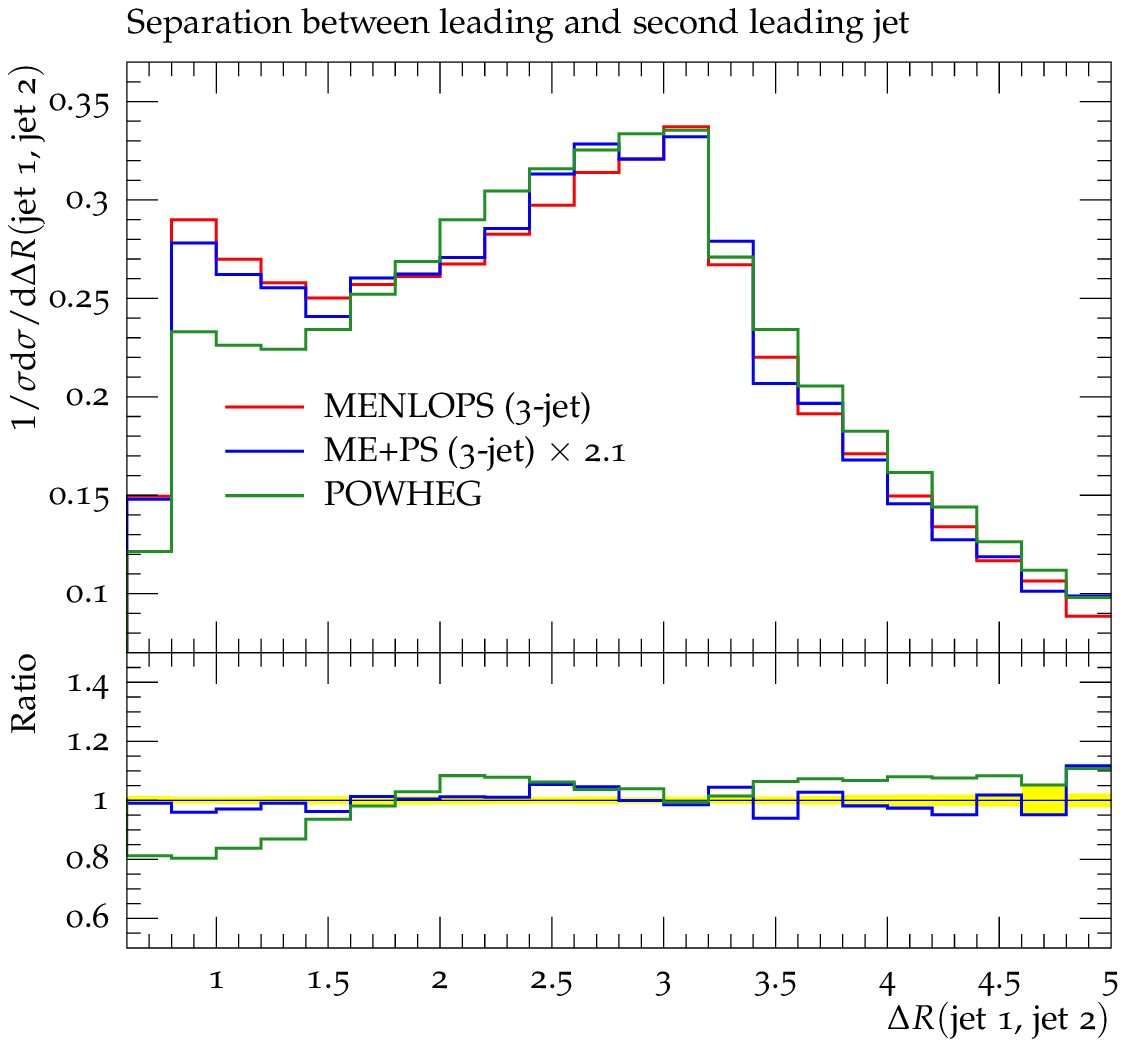}\vspace*{-5mm}
  \end{center}
  \caption{Left: The transverse momentum of the reconstructed Higgs boson 
  in the gluon fusion process at nominal LHC energies (14 TeV).
  Right: Separation in $\eta$-$\phi$ space of the first and second hardest 
  jet in Higgs-boson production via gluon fusion at nominal LHC energies.
  \label{Fig:ggh_hpt_drjj}}
\end{figure}

\begin{figure}[p]
  \begin{center}
    \includegraphics[scale=0.5725]{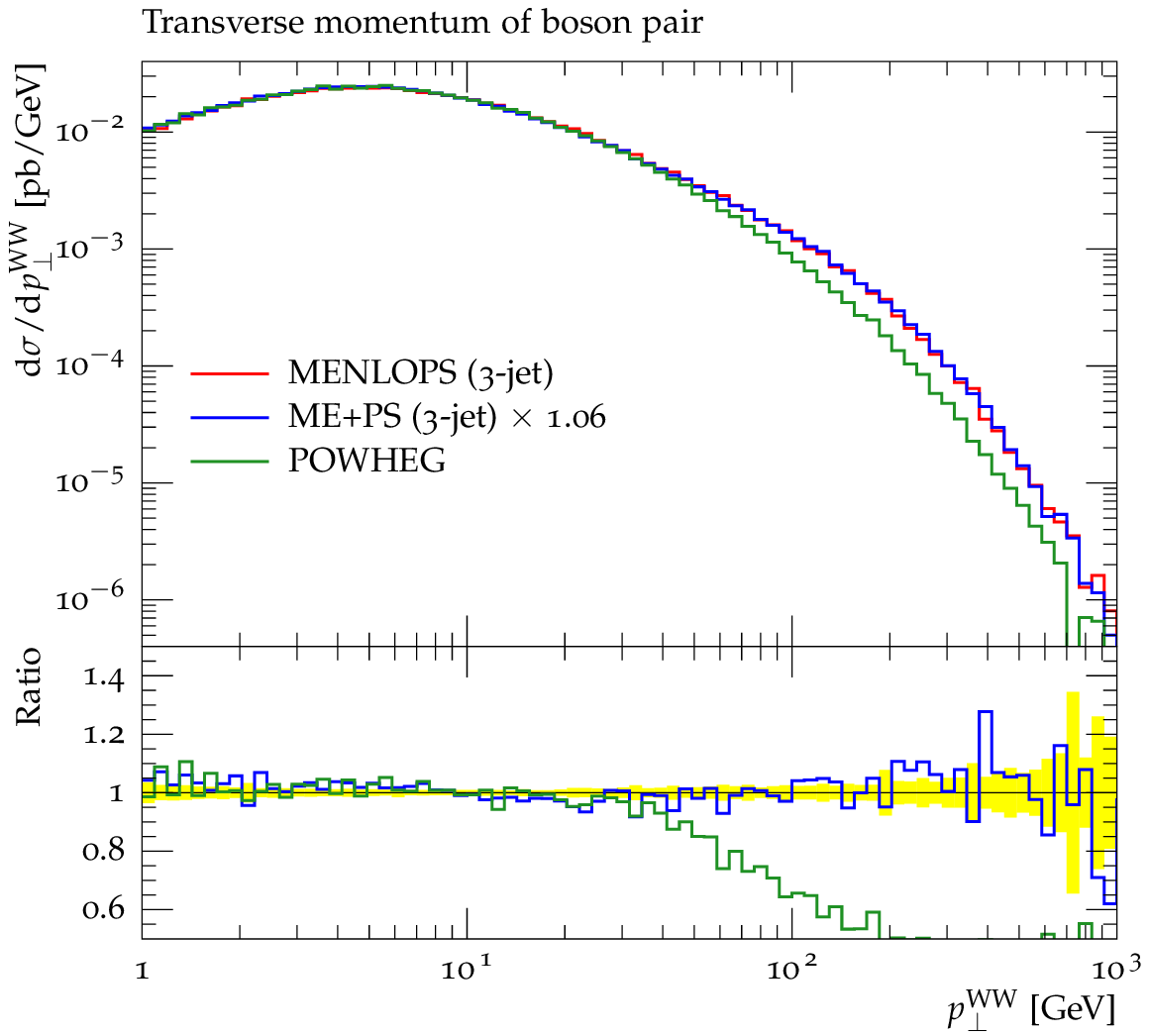}\hfill
    \includegraphics[scale=0.575]{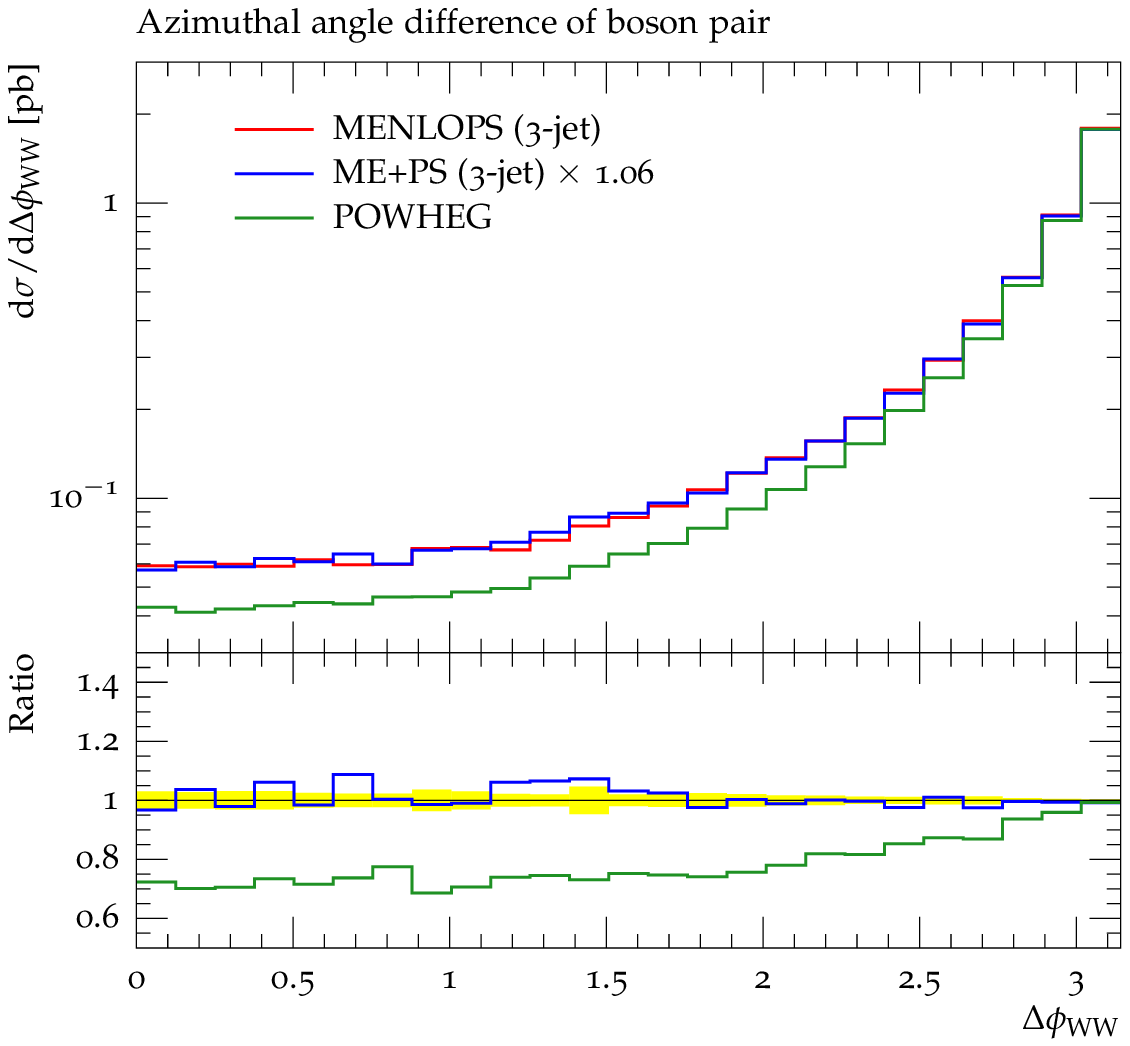}\vspace*{-5mm}
  \end{center}
  \caption{Left: The transverse momentum of the reconstructed $W^+W^-$ system
  in $W^+[\to\! e^+\nu_e]\;W^-[\to\! \mu^-\bar\nu_\mu]$ events at nominal 
  LHC energies (14 TeV). Right: Separation in $\eta$-$\phi$ space of the first
  and second hardest jet in $W^+W^-$ production at nominal LHC energies.
  \label{Fig:ww_wwpt_drjj}}
\end{figure}

We finally present some predictions for the production of Higgs-bosons through 
gluon-gluon fusion and for the production of $W^+[\to\! e^+\nu_e]\;W^-[\to\! \mu^-\bar\nu_\mu]$
at nominal LHC energies ($\sqrt{s}=$14 TeV). Virtual matrix elements for these
analyses have been taken from~\cite{Dawson:1990zj} and~\cite{Dixon:1998py}, respectively.
Results are shown in Figs.~\ref{Fig:ggh_hpt_drjj} and~\ref{Fig:ww_wwpt_drjj}. 
We observe very small uncertainties related to the intrinsic parameters of the 
MENLOPS approach. A detailed discussion is found in~\cite{Hoeche:2010kg}.

\section*{Acknowledgements}
SH acknowledges funding by the SNF (contract 200020-126691) and by the 
University of Zurich (FK 57183003).  MS and FS gratefully acknowledge 
financial support by MCnet (contract MRTN-CT-2006-035606). MS further 
acknowledges financial support by HEPTOOLS (contract MRTN-CT-2006-035505) 
and funding by the DFG Graduate College 1504. FK and MS would like to thank 
the theory group at CERN and the IPPP Durham, respectively, for their kind 
hospitality during various stages of this project.

\begin{footnotesize}

\end{footnotesize}

\vfill

\vspace*{1cm}


\end{document}